\documentclass[graphics, twocolumn, usenatbib]{mn2e}
\usepackage{natbib} 
\usepackage{epsfig} 
\usepackage{graphicx} 
\usepackage{color}
\usepackage{aas_macros} 
\usepackage{amssymb}
\usepackage{amsmath}
\usepackage[title]{appendix}

\newcommand{\lya}{Lyman-$\alpha$~}

\newcommand{\msolar} {$\rm{M_{\odot}}~$}
\newcommand{\msolarc} {$\rm{M_{\odot}}$}

\newcommand{\zsolarc} {$\rm{Z_{\odot}}$}
\newcommand{\molH} {$\rm{H_2}$~}
\newcommand{\molHc} {$\rm{H_2}$~}

\newcommand{\inten} {$\rm{ erg\ cm^{-2}\ s^{-1}\ Hz^{-1}\ sr^{-1}}$~}

\newcommand{\JJ} {\rm{$J_{21}$}~}
\newcommand{\JJc} {\rm{$J_{21}$}}

\newcommand{\msun}{~\mathrm{M}_{\odot}}
\newcommand{\zsun}{~\mathrm{Z}_{\odot}}

\def\etal{{\it et al.}~}
\begin{document}
\title[Analytic resolution to LW vs. metals]{An analytic resolution to the competition between Lyman--Werner radiation and metal winds in direct collapse black hole hosts}
\author[B. Agarwal \etal] 
{Bhaskar Agarwal$^{1}$\thanks{E-mail: bhaskar.agarwal@uni-heidelberg.de}, John Regan$^{2, 1}$, Ralf S. Klessen$^{1}$, Turlough P. Downes$^{2}$ \&
\newauthor{Erik Zackrisson$^3$} \\
$^1$Universit{\"a}t Heidelberg, Zentrum fur Astronomie, Institut fur Theoretische Astrophysik, Albert-Ueberle-Str. 2, D-69120 Heidelberg \\
$^2$Centre for Astrophysics \& Relativity, School of Mathematical Sciences, Dublin City University, Glasnevin, D09 Y5N0, Dublin, Ireland \\
$^3$Department of Physics and Astronomy, Uppsala University, Box 515, SE-751 20 Uppsala, Sweden\\
}


\maketitle

\begin{abstract} 

  \noindent A near pristine atomic cooling halo close to a star forming galaxy offers a natural pathway
  for forming massive direct collapse black hole (DCBH) seeds which could be the progenitors of
  the $z>6$ redshift quasars. The close proximity of the haloes enables a sufficient Lyman-Werner
  flux to effectively dissociate \molHc in the core of the atomic cooling halo. A mild background
  may also be required to delay star formation in the atomic cooling halo, often attributed to distant background galaxies.
  In this \textit{letter} we investigate the impact of metal pollution
  from both the background galaxies and the close star forming galaxy under extremely unfavourable conditions 
  such as instantaneous metal mixing. We find that within the 
  time window of DCBH formation, the level of pollution never exceeds the critical threshold 
  (Z$_{cr} \sim 1 \times 10^{-5}$ \zsolarc), and attains a maximum metallicity of 
  Z  $\sim 2 \times 10^{-6} \ \rm Z_{\odot}$. As the system evolves, the metallicity eventually exceeds 
  the critical threshold, long after the DCBH has formed.

\end{abstract}

\begin{keywords}
Cosmology: theory -- large-scale structure -- first stars, methods: numerical 
\end{keywords}


\section{Introduction} \label{Sec:Introduction}
\begin{figure}
  \centering 
  \begin{center}
      \centerline{
        \includegraphics[width=8cm]{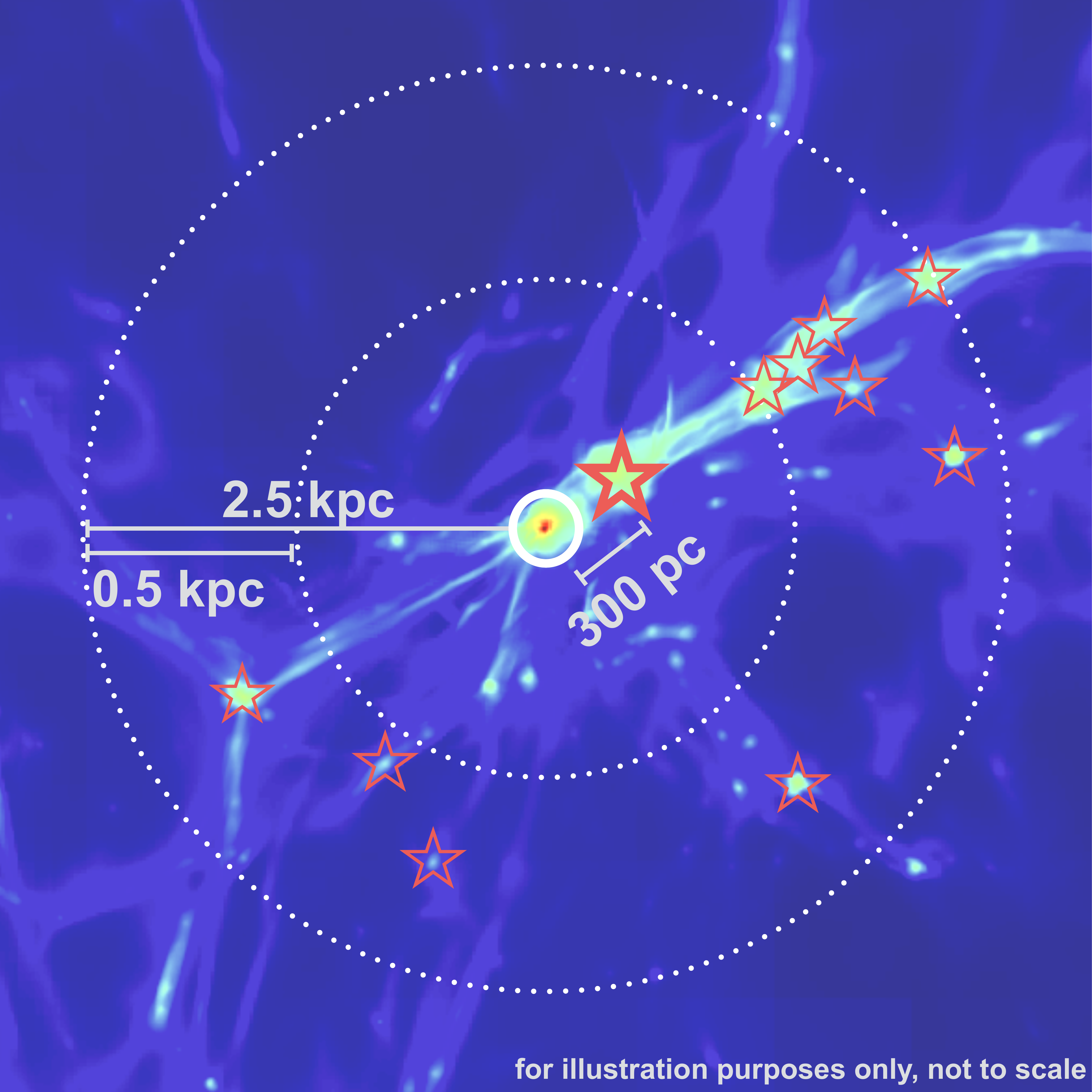}}
        \caption[]
         {\label{cartoon} This is the model we investigate. Two synchronised proto galaxies
           sit in a clustered region exposed to a background LW radiation field.
           The DCBH halo is centred within the small solid circle. The neighbouring halo is denoted
           by the large ``star'' immediately to the right of the DCBH halo.
           We investigate the impact of metal pollution from the galaxies (marked
           as red stars) on both of the (synchronised) haloes growing at the centre. The background
           galaxies must provide a sufficient LW background to delay the collapse of the
           central halos but crucially not pollute the two synchronised galaxies with metals.
        }
      \end{center}
  \end{figure}

{The discovery of a significant number of quasars at $z>6$ hosting 
massive black holes with masses exceeding a billion solar masses \citep{Fan_2006b, Mortlock_2011, 
Venemans_2013, Wu_2015} has challenged our understanding of how super-massive black holes (SMBHs)
 form in the first billion years of our Universe's evolution.}
{Three main avenues have emerged to explain their formation. Firstly Population III (PopIII) remnants
could act as the seeds of these black holes \citep[e.g.][]{Madau_2001, Bromm_2002, Bromm_2003,
  Milosavljevic_2009}. Secondly, the seeds may themselves be massive, $10^{4-5}\msun$, 
  and form as a result of the collapse of objects with masses significantly larger than
typical PopIII remnants \citep[e.g.][]{Loeb_1994, Koushiappas_2004, Begelman_2006, Wise_2008a, Regan_2009b}.
Finally, the formation of massive black hole seeds could result from collisions in a stellar cluster
\citep[e.g.][]{Begelman_78a, Devecchi_2008, Katz_2015, Yajima_2016}, or due to high inflow rates in the central region
resulting from massive galactic collisions in the early Universe \citep{Mayer_2010, Mayer_2014}.} \\
\indent In this study we examine the second avenue outlined above, the so-called direct collapse (DC) mechanism.
The DC mechanism is thought to occur when a halo
is able to grow to the atomic cooling threshold, i.e. virial temperature T$_{\rm{vir}} > 10^4$ K,
without forming stars. This can be achieved through the destruction of \molH in the halo either
through a background radiation field \citep{Machacek:2001p150,Oh_2002}  or also through the impact of
relative streaming velocities \citep[Hirano et al. in prep; Schauer et al. in prep]{Tseliakhovich_2010, Tanaka_2014}. Once
the halo reaches the atomic cooling limit, \lya cooling becomes effective and the
halo collapses isothermally at a temperature, T $\sim 8000$ K, leading to the formation of a
$10^{4-5} \msun$ direct collapse black hole (DCBH).
Furthermore, the halo must also avoid
significant metal pollution, in order to avoid fragmentation \citep[e.g.][]{Clark_2008}. \\
\indent In this letter we study a specific example extracted from the recent simulations of
\cite{Regan_2017} (hereafter R17). 
They modelled a scenario where a background LW radiation field
is created by a cluster of nearby galaxies, termed \textit{background galaxies}, surrounding
two haloes \citep{Dijkstra_2008, Agarwal_2014b,Visbal_2014b}. One of these haloes is the DCBH candidate,
termed the \textit{target halo}. The background LW intensity required to delay the collapse of the two haloes is
J$_{bg}\sim 100$\footnote{Previous studies have reported that a $100-1000$ times smaller value of J$_{bg}$ is sufficient
  to suppress Pop III SF in similar mass haloes \citep{Machacek:2001p150,Yoshida:2003p51,OShea:2008p41}. We attribute
  the difference to the fact that simulations of R17 extract haloes from \textit{rare-peaks}, which was not the case
  in the aforementioned studies and that the delay required for synchronisation is longer. } in units of J$_{21}$, i.e.
$10^{-21}$ \inten.  R17 find that this J$_{bg}$ is not sufficient to prevent \molH formation in the core of the central haloes. For the complete
destruction of \molH throughout the target halo, one of haloes, the \textit{neighbour}, must form
stars (see Figure \ref{cartoon} for an illustration) shortly before the target halo undergoes
runaway collapse - this window is the synchronisation time. The rapid star formation in the neighbour produces an intense burst
of radiation which completely prevents \molH formation in the core of the target halo pushing it onto the isothermal cooling track and towards DCBH formation.

R17 show that in this scenario the deleterious effects of
photo-evaporation from the neighbour are avoided. 
However, the treatment of R17 neglected the impact of metal pollution from
both the background galaxies and the neighbour. Here, using the semi-analytic
model developed by \cite{Agarwal_2017} (hereafter A17) we investigate the impact of metal
pollution from both the background galaxies and the neighbour. For the purposes of gaining
the most insight into metal pollution we assume a \textit{reductio ad absurdum} approach, where the 
parameters chosen in this study are most unfavourable for DCBH formation. 
In particular, we assume instantaneous metal mixing and that the metals
are ejected from the background galaxies as soon as they become star forming.\\
\indent In \S \ref{sync_model} we outline the direct collapse formation model that we explore and
discuss both the radiation field and metal field expected in such a model. In \S \ref{results} we outline
our results and finally in \S \ref{conclusions} we present our conclusions.

\section{Working Model} \label{sync_model}

The model described below builds on the existing framework of R17 for initial inputs for the
synchronous halo pairs from their simulation(s), and on A17 for computing the metallicity of the target
halo. 
\subsection{Background Radiation Field} \label{bg_field}
\begin{figure}
  \centering 
  \begin{center}
      \centerline{
        \includegraphics[width=\columnwidth]{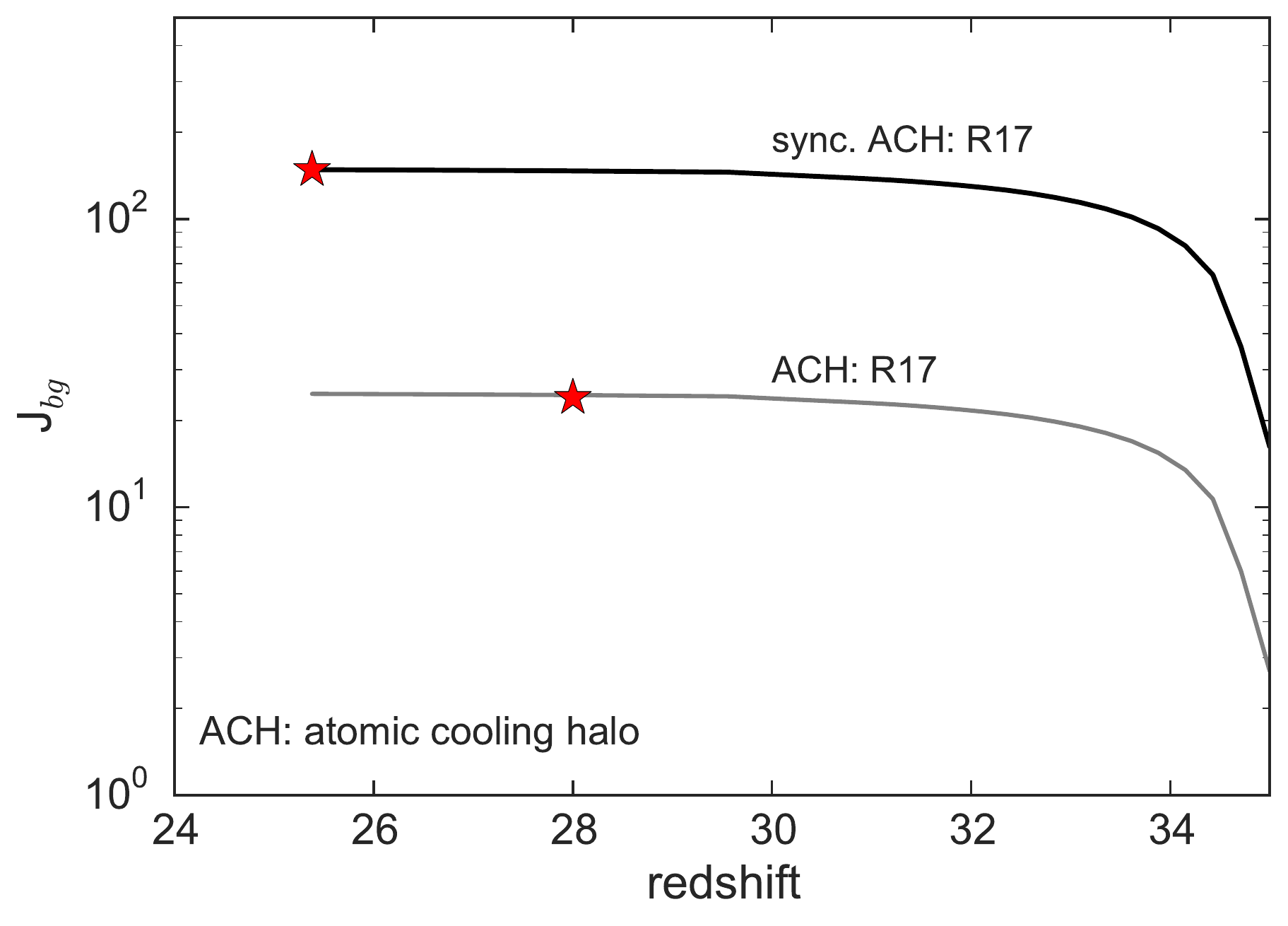}}
        \caption[]
                {\label{JWFlux} The evolution of the LW field as function of redshift. The field
                  is turned on at redshift z = 35, corresponding to onset of SF in the background galaxies. The red star 
                  marks the epoch where SF occurs, in the absence of a nearby irradiating source. 
                  The thick solid line is LW intensity produced by
                  all of the background galaxies required to delay the collapse sufficiently till $z\sim 25.4$
                   to allow for synchronised DC as per R17. The thin solid line is the background in R17 that produces 
                  an atomic cooling halo, which undergoes SF at $z\sim 28$, i.e. before the neighbouring galaxy becomes SF.                  
        }
      \end{center}
  \end{figure}

The required background LW radiation field, as found in R17, to allow both the target halo and the 
neighbour halo to grow sufficiently is J$_{\rm{LW}} \gtrsim 100$ \JJc.
In order to calculate the stellar mass required to create the necessary LW intensity {we use SEDs
derived from \cite{2010A&A...523A..64R} rescaled to a Kroupa \citep{Kroupa2001} IMF}. We assume the stellar populations have a metallicity of $Z\sim 5\times10^{-6}\ \rm Z_{\odot}$, as they are expected to form in very low
metallicity gas and therefore produce copious amounts of LW radiation. We turn the background
galaxies on at a redshift z = 35 as was done in R17. We assume a constant star formation rate (SFR)
over a 60 Myrs period (from z $\sim 35$ up until z $\sim 25$). In order to produce a constant LW
intensity of J$_{\rm{LW}} \gtrsim 100$ \JJ a final stellar mass of M$_{\star,bg}^{tot} = 5 \times 10^6$
\msolar is required, {within the sphere of radius $\sim 2.5$ kpc around the target halo}. 
The background galaxies are assumed to be made up of a total of $n_s$ sub-systems which
together provide the cumulative intensity required. The model is outlined for illustrative purposes
in Figure \ref{cartoon}. The value of $n_s$ has no impact on our calculations which depend only on
the SED assumed for the stellar population we now describe. The value of the LW intensity
can be computed for each subsystem as  \citep{Agarwal_2012}
\begin{equation}
  J_{bg,sub}(t_i) = {\dot{E}_{LW}(t_i) \over 4\pi^2 D^2} {M_{\star ,6}(t_i) \over \Delta \nu J_{21}}
\end{equation}
where $\dot{E}_{LW}(t_i)$ is the LW emission (erg/s) for a given age of a $10^6 \msun$ stellar
population at a given timestep $t_i$, $\Delta\nu$ is the difference between the highest and the
lowest frequency in the LW band, D is the distance of each sub-system from the DCBH halo,
$M_{\star,6}$ is the mass of each sub-system normalised to $10^6\msun$, and \JJ is the normalisation
factor for the specific intensity. The extra factor of $\pi$ in the denominator accounts for the
solid angle. We then simply compute the total background at each redshift as $n_s\mathrm{J}_{bg,sub}$.
The average distance between the sub-systems and the target galaxy is set at 2.25 kpc.\\
\indent Figure \ref{JWFlux} shows  the J$_{bg}$ used here as a function of redshift, z (thick solid line). The LW intensity increases
as the stellar mass increases reaching a value of J$_{bg} \sim 100$ \JJ at z $\sim 31$ - this is the
minimum background intensity required in the models of R17, and we take this as our fiducial
case\footnote{Throughout we take z2540\_100\_250 in R17 as the fiducial case.}. 
We assume that the 10 background galaxies all become active at approximately z = 35 with an initial mass of
M$_{*} \sim 8 \times 10^4 \msun$. Over the redshift range z = 35 to z = 25.4 the mass of each background 
galaxy grows with a constant SFR of $0.01\rm \msun/yr$. This results in the total stellar mass over all subsystems 
to grow from M$_{*} \sim 8 \times 10^4$ \msolar to M$_{*} = 5 \times 10^6$ \msolarc. It is this total stellar mass, aged accordingly,
 that produces the required J$_{bg}$ and can be distributed among any number of subsystems.


The goal of this study is to test if the background galaxies pollute the synchronised pair. The metallicity of the
pair is linked to its separation from the background galaxies and to their stellar mass. If the background galaxies are too
close they will inevitably pollute the environment of synchronised haloes over the timescale of T $\sim 60 $ Myr for which they must be active, while if they are too distant the LW intensity will be insufficient.

\subsection{Near Neighbour Radiation Field}

The synchronised pair must be at a mutual separation of d $\lesssim 300$ pc for a stellar mass of
M$_{*, burst} = 10^5\msun$. The SFR assumed for the neighbouring galaxy is set to $0.1 \rm \msun/yr$, and the burst itself lasts for 1 Myr. The neighbour attains M$_{*, burst}$ at $z\sim25.4$, with the DC
in the target halo occurring at $z\sim24.2$, consistent with the case$^2$ of R17 which is taken here as the fiducial model. For an assumed separation of $d \sim 276$~pc, the neighbour
provides a LW specific intensity of $\sim 1000 \rm \ J_{21}$, which completely destroys \molH within the target halo (see R17 Fig. 2). We therefore also examine the impact of metal pollution
from the neighbour bearing in mind that the time for which the neighbour is ``on'' is of
the order of T$_{\rm{on}} \sim 9$ Myr and it would also take at least 2 Myr (corresponding to the lifetime of a $100 \msun$ star) for the
metals to be expelled from supernovae explosions after star formation begins.
\begin{figure*}
  \centering 
  \begin{minipage}{175mm}      \begin{center}
      \centerline{
        \includegraphics[width=14cm]{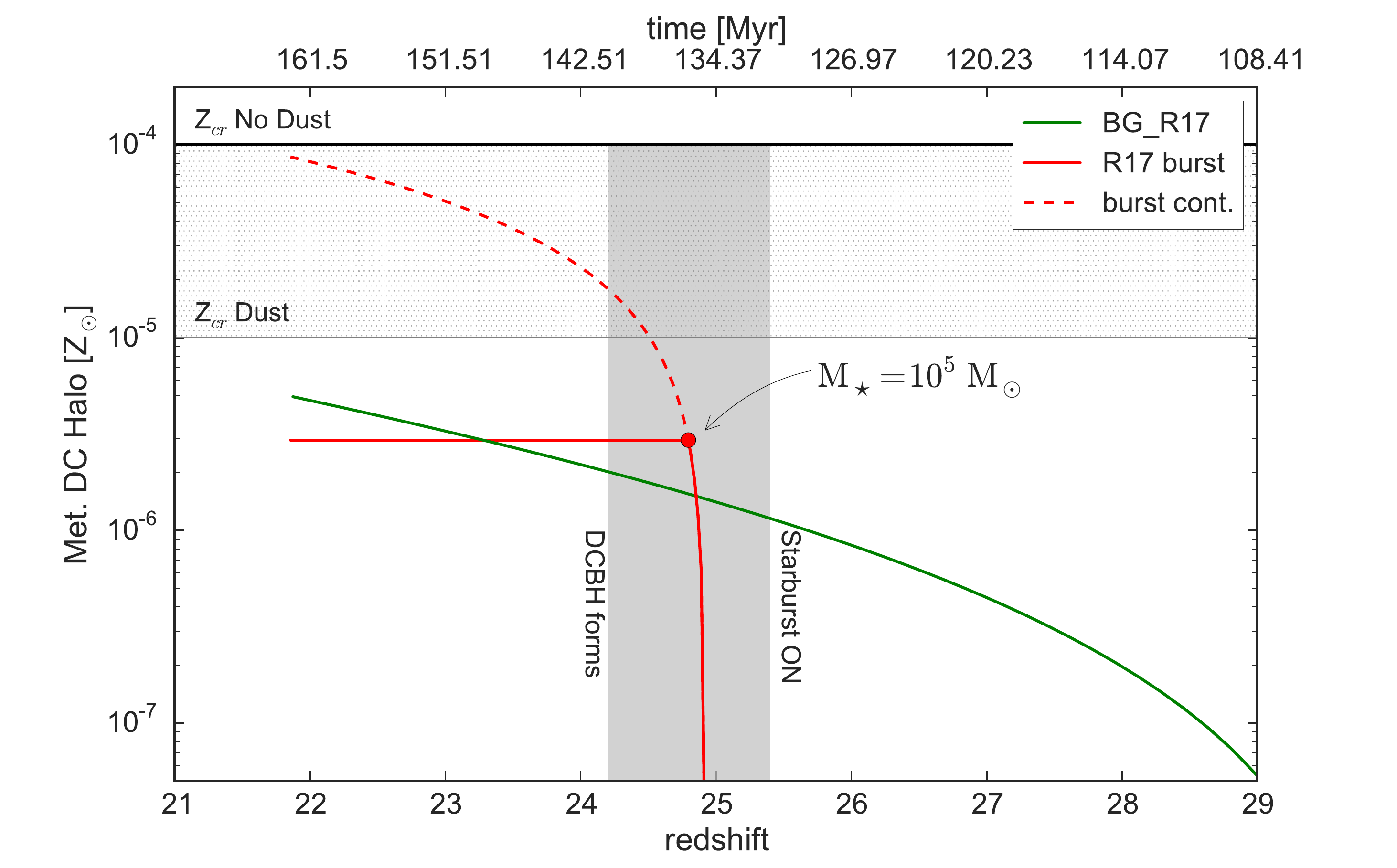}}
      \caption{Metallicity evolution of the target halo due to metals from the background galaxies and the nearby source.
        Solid lines indicate the fiducial case considered in this work consistent with one of the simulation runs of R17.
        In grey we show the time window for DCBH formation which is $\sim 9$ Myr between the nearby source turning on at $z=25.4$,
        and the DCBH forming in the target halo at $z=24.2$.  The red dot indicates the epoch of metal pollution corresponding to
        the nearby source attaining a stellar mass  of $10^5 \msun$, after which no further SF is permitted. The dashed red line
        indicates the metallicity evolution of the target halo, assuming that the nearby source continues to form stars after this
        epoch. The green solid line indicates the metallicity of the target halo due to the background LW field as seen in R17 that
        produces a synchronous pair of atomic cooling haloes at $z\sim 25$. 
        }
        {\label{money_plot}
          
        }
      \end{center} \end{minipage}
  \end{figure*}


\subsection{Metal Pollution Modelling} \label{metals}
Metal pollution of the target haloes is computed following the method presented in A17.
In order to model this process due to the surrounding galaxies we make some simplifying
assumptions regarding both the star formation efficiency and the mass outflow rates from these
systems. The mass loading factor is defined as $\eta = {{\dot{\rm{M}}_{\rm{outflow}}} / {\dot{\rm{M}}_{*}}}$. 
Here $\dot{\rm{M}}_{\rm{outflow}}$ is the mass outflow rate and  $\dot{\rm{M}}_{*}$ is the SFR. Owing to the small masses of our haloes ($<10^8 \msun$) we set $\eta$ = 20 \citep[Dalla Vecchia in prep.]{Muratov2015}. 
Given the stellar masses that lead to the required LW intensity as a function of redshift, we must
now also compute the metal pollution of the target halo. To calculate this we first need to compute
the fraction of metals and outflow from each galaxy which intersects with the target halo,
\begin{align}
  \rm{M}_{\rm{inter, out}}   &= \rm{M}_{\rm{outflow}} * \mathit{f} \\
  \rm{M}_{\rm{inter, metals}} &= \rm{M}_{\rm{metals}} * \mathit{f},
\end{align}
where the mass in metals is computed as $\rm M_{metals} = \mathit{y}M_{\star}$. We define
$y= 0.032$ as the metal yield factor for a Kroupa type IMF (A17). The intersection term $f$ for the target halo is defined as
\begin{equation}
  f =min\left(0.5, {\pi R_{bind}^2 \over 4 \pi D^2}\right),
\end{equation}
where $\rm{R_{bind}} = GM_{target}/v_{wind}^{2}$ is the gravitational binding radius of the target halo and $D$ is the
average separation between the background galaxies and the target halo. The target halo is assumed
to have a constant total mass growth rate starting from $4.3 \times 10^5\msun$ at $z=35$ to
$8 \times 10^6\msun$ at $z=25$, corresponding to a virial temperature of $T_{vir} = 2000$ K and  $T_{vir} = 10^4\ \rm K$
respectively. Thus the resultant metallicity of the target halo at any given redshift $i$ then
becomes (A17) 
\begin{equation}
Z^{i+\Delta} = {\sum\limits_{z=35}^{i} {\rm{M}_{inter, metals}} \over {\rm{M}_{baryons} + \sum\limits_{z=35}^{i} \rm{M}_{inter, out}}} \ { 1 \over Z_{\odot}}, 
\end{equation}
where $\Delta$ is the time delay for the winds to reach the target halo with a velocity of
$v_{wind}=100\ \rm km/s$. For the background galaxies $\Delta = {D}/{v_{wind}}\sim 25 \rm \ Myr$
for $D=2.25 \ \rm kpc$. The metallicity of the target halo due to the nearby neighbour is computed
in a similar manner but with the stellar masses and separations updated accordingly. Given the close
proximity and the relatively short timescale, as compared to the background galaxies, an additional
delay of $t_{SN}$ is also added to the time delay for the nearby source,
i.e. $\Delta = t_{SN} + ({d}/{v_{wind}}) \sim 5\ \rm Myr$, where $t_{SN}$ is the supernova timescale.
We assume that metal mixing is efficient, instantaneous and uniform once it reaches the target halo.

\section{Results and Discussion} \label{results}
Given the presence of a sufficient level of external LW flux, a DCBH may form in an atomic cooling
halo comprising of metal--poor gas, as long as the metallicity of the gas is less than a critical
value. This critical value is found to be $Z_{cr}\sim 10^{-4}\zsun$ for dust--free,
and $Z_{cr}^{dust}\sim 10^{-5}\zsun$ for dust--rich environments \citep{Omukai_2008, Latif_2016}.\\
\indent We plot the evolution of the target halo's metallicity{, Z$_t$}, in Fig.~\ref{money_plot} where the
solid lines depict our fiducial case consistent with R17. {The grey region marks the time window of 
DCBH formation, where the nearby source turns on at $z=25.4$, and the target halo forms
a DCBH at $z=24.2$. This marks the time window where the LW radiation from the neighbour is required
to completely destroy H$_2$ and facilitate atomic H cooling in the target halo. If the metallicity of the target halo
exceeds the $Z_{cr}$ in this time frame, then no DCBH formation can occur in the target halo.
Note that this is one of the cases of R17 with the longest time delay
($\sim 9$ Myr) between the source being turned on and a DCBH forming in the target halo. We chose 
this particular case to maximise the possibility of polluting the target halo, thus studying the effect of metal pollution
at its highest efficiency.} 
The metallicity due to the nearby source remains constant after a stellar mass of
$10^5 \msun$ is attained, as no further SF is permitted in the nearby source, consistent with R17.
The metallicity due to the outflow from background galaxies evolves depending on their star
formation history, as discussed in the previous section. The curves are plotted at the appropriate
redshift, after taking into account the time delay for the winds to reach the target halo which is
$\Delta = 25$ and 5 Myr from the background galaxies and nearby source respectively. 
{We find that for our fiducial case (solid curves), metal pollution from both the background galaxies and the nearby
source never exceeds the critical metallicity and $Z_t<Z_{cr}^{dust}<Z_{cr}$, where $Z_t$ is the target halo metallicity.
The maximum metallicity of the target halo, $Z_t\sim 3\times10^{-6}\zsun$, is attained
 at the time when the DCBH forms. Even if the nearby source is allowed to continue its
 starburst after a stellar mass of $M_{\star} = 10^5 \msun$ is reached\footnote{Feedback from a growing PopIII stellar population is expected to prevent PopIII galaxies from growing to larger sizes \citep{Xu_2013}}, the target halo
 maintains $Z_t<Z_{cr}$ in the DCBH time window. Even with a background of J$_{BG} \sim 1000$,
 metal pollution is inefficient and is not able to prevent DCBH formation in the target halo. A lower value of 
J$_{bg}$, which would require lower stellar masses, would only bring down the metallicity due to
background galaxies in the target halo (see \S \ref{metals}). Thus, lowering the J$_{bg}$ would only further strengthen our result.} 
Note that we have assumed instantaneous mixing of the
metals in the target halo, while in reality, this would hardly be the case
\citep{Cen_2008, Smith_2015}. Any additional delay due to the mixing timescale of metals would
increase $\Delta$, thereby shifting the curves further to the left and reducing the overall
metallicity of the target halo in the DCBH time window. {For example, the sound crossing time for
our target halo is of the order of t$_s \sim 15$ Myrs.}

We have extracted a worst case scenario (i.e. a case with the longest time taken for DCBH to form) from the
framework of R17, and further applied a \textit{reductio ad absurdum} approach to allow maximal
metal pollution of the DCBH target halo from the LW radiation sources. Our results indicate that
metal pollution of a possible DCBH host due to background galaxies, or the nearby irradiating source
is insufficient in raising its metallicity to values where fragmentation into stars occurs. 

\section{Conclusions} \label{conclusions}
We have investigated here the impact of metal pollution from both a cluster of
background galaxies and the nearby neighbour galaxy as a
mechanism for DCBH formation. The Lyman-Werner radiation from the cluster of background galaxies
suppresses PopIII formation in the two haloes enabling them to evolve until one eventually becomes star forming
 and provides the necessary LW radiation field to the target halo where DCBH formation {can occur}.
 Our semi-analytical model of metal pollution, shows
that the metallicity of the target halo halo remains well below
the critical threshold during the entire time window for DCBH formation. After the DCBH forms, metals
from both the background galaxies and the neighbouring galaxy will continue to pollute the DCBH halo as the system evolves.

\section*{Acknowledgements}
The authors would like to thank Jarrett L. Johnson for his useful comments on the manuscript, and Eric Pellegrini and Simon Glover for helpful discussions. BA and RSK acknowledge the funding from the European Research Council under the European Community's Seventh Framework Programme (FP7/2007-2013) via the ERC Advanced Grant STARLIGHT (project number 339177).
Financial support for this work was also provided by the Deutsche Forschungsgemeinschaft via SFB 881, "The Milky Way System" (sub-projects B1, B2 and B8) and SPP 1573, "Physics of the Interstellar Medium" (grant number GL 668/2-1). RSK acknowledges the Universit\"{a}t Heidelberg, Interdiszipli\"{a}res Zentrum f\"{u}r Wissenschaftliches Rechnen, Im Neuenheimer Feld 205, 69120 Heidelberg, Germany. J.A.R. acknowledges the support of the EU Commission through the
Marie Sk\l{}odowska-Curie Grant - "SMARTSTARS" - grant number 699941.

\bibliographystyle{mn2e_w}
\bibliography{./BIBTEX/mybib}
\end{document}